\documentclass[12pt,tightenlines,nofootinbib]{revtex4}
\usepackage{amssymb}
\usepackage{amsmath}
\usepackage{amsfonts}
\usepackage[stable]{footmisc}
\usepackage{hyperref,blkcntrl,moredefs,relsize,attrib,tikz,booktabs,capt-of,float,varwidth,epsdice,graphicx}

\newcommand{\qed}{\nobreak \ifvmode \relax \else
      \ifdim\lastskip<1.5em \hskip-\lastskip
      \hskip1.5em plus0em minus0.5em \fi \nobreak
      \vrule height0.75em width0.5em depth0.25em\fi}

\newcommand{\ket}[1]{\left| #1 \right>} % for Dirac bras
\let\baraccent=\= % rename builtin command \= to \baraccent
\renewcommand{\=}[1]{\stackrel{#1}{=}} % for putting numbers above =

\begin{document}

\title{Bell's Theory of Beables and the Concept of `Universe'}

\author{Ian T. Durham}
\email[]{idurham@anselm.edu}
\affiliation{Department of Physics, Saint Anselm College, Manchester, NH 03102}
\date{\today}

\begin{abstract}
From its earliest days nearly a century ago, quantum mechanics has proven itself to be a tremendously accurate yet intellectually unsatisfying theory to many. Not the least of its problems is that it is a theory about the results of measurements. As John Bell once said in introducing the concept of `beables', it should be possible to say what \textit{is} rather than merely what \textit{is observed}. In this essay I consider the question of whether a universe can be a (nonlocal) beable and what that implies about the fundamental nature of that universe. I conclude that a universe that is a beable within the framework of some theory, cannot also be fundamental.
\end{abstract}

\maketitle

\section{`Beables' and induction}
When I was in graduate school in Scotland, I was told the following parable by my advisors. An economist, a mathematician, and a logician were on a train traveling north. Just after they passed the Scottish border they noticed a single cow standing in a field. The economist remarked, ``That cow is brown. All cows in Scotland must be brown.'' The mathematician replied, ``No, \textit{one} cow in Scotland is brown.'' The logician quietly but firmly muttered ``No, one \textit{side} of one cow in Scotland is brown.'' There are many versions of this parable involving a variety of professions and there are any number of lessons to be taken from it. It is usually meant as a dig at one of the particular professions that is included, especially when told by a member of one of the other professions. At the heart of the parable, though, is an open question: how much can we reasonably infer from a given observation?

It is worth noting here that my advisors were both mathematicians. As such, I always had the impression that the parable, as they told it, was meant as a dig at both economists \textit{and} logicians. Clearly the economist has over-extrapolated from the given data. That point is hardly up for debate. But has the logician \textit{under}-extrapolated? The fact is that our intrepid travelers do not know if the cow is the same color on both sides absent additional information. It is entirely possible that the mathematician, in casually suggesting that the cow was entirely brown, was wrong. Yet, in our own experience with cows, most of us would probably think it highly unlikely that a cow had such asymmetrical coloring as to be entirely different on either side.

In our effort to understand the world we inhabit, we wrestle with such questions of inference as a matter of course. At the heart of the problem of inference, more properly known as the problem of induction~\cite{Popper:1959aa}, is what John Bell referred to as the `subject-object distinction'\cite{Bell:2004aa}. This distinction is best understood in the context of quantum mechanics but it is not limited to that realm. Quantum mechanics is ostensibly a theory about the results of measurements. Measurements are performed `on' systems (object) and presuppose that something or someone (subject) must be doing the measuring. But as Bell pointed out, precisely \textit{where} or \textit{when} to draw a distinction between subject and object is not manifest in the theory itself. This inherent ambiguity continues to be the source of much debate.

As humans, we naturally tend to anthropomorphize. The very word `measurement' suggests a human-centric outlook. So it is that, for many, the subject-object distinction is interpreted as concerning knowledge. By making a measurement on a system (object), the measurer (subject) has acquired knowledge about that system. Bell finds this unsatisfying. He suggests that any accurate, final theory of physics (should one ever be found) could not be about the acquisition of knowledge.
\begin{quote}
[It] could not be fundamentally about `measurements', for that would again imply incompleteness of the system and unanalyzed interventions from outside. Rather it should again become possible to say of a system not that such and such may be \textit{observed} to be so but that such and such \textit{be} so~\cite{Bell:2004aa}, p. 41.
\end{quote}
Rather than being about \textit{observ}ables, such a theory would need to be about \textit{be}ables.

On its face this appears to be a bold prescription. Presumably any such final theory of physics would provide us with a means of obtaining complete knowledge of the world. But it's not clear that objectively complete knowledge of the world is even attainable in theory let alone in practice. Bell is more practical. He recognizes that any final theory would need to somehow clarify or circumvent the ambiguities in the subject-object distinction that arise in any of our existing theories, quantum mechanics in particular. Universal beables may not be knowable, but local ones, as in those bounded within a particular region of space-time, might. It is only by first understanding local beables that we might have some hope of constructing a final theory.

It is worth noting exactly what Bell means by `beable'. He actually initially uses the word in two slightly different contexts. In the first context he suggests that beables within a given theory must be describable in classical terms since ``they are there''~\cite{Bell:2004ab}, p. 51. Here he (oddly) seems to be motivated by Bohr, saying 
\begin{quote}
[b]y `classical terms' here Bohr is not of course invoking particular nineteenth century theories, but refers simply to the familiar language of everyday affairs, including laboratory procedures, in which objective properties -- \textit{beables} -- are assigned to objects~\cite{Bell:2004aa}, p. 41.
\end{quote}
Such beables, he notes, must necessarily include things like the settings of switches and knobs, and the readings of instruments.

In the second context in which he initially employs the term, he suggests that beables in a given theory are expressly physical quantities in the sense ``familiar already from classical theory''~\cite{Bell:2004ab}, p. 52. The example he cites in order to clarify this point is the contrast between the \textbf{E} and \textbf{H} fields in electromagnetism, which he suggests are physical, and the \textbf{A} and $\phi$ potentials, which are not. Make no mistake---Bell explicitly says that \textbf{E} and \textbf{H} are beables within the context of Maxwell's electromagnetic theory: ``the fields \textbf{E} and \textbf{H} are `physical' (beables we will say)...''~\cite{Bell:2004ab}, p. 52.

In both of these contexts, the beables form the ontology of the theory. It's actually worth asking what we mean by this. All physical theories are `about' something. One might say that the beables are what a theory is about. So Maxwell's electromagnetic theory is about electric and magnetic fields and so those fields are beables within that theory. But that doesn't quite capture the meaning implied in the first context where the beables are said to be objective properties, including instrument settings, that are applied to objects. In classical electromagnetic theory, we are accustomed to thinking of electric and magnetic fields as having their source in charged particles. Thus, the fields are objective properties of the charged particles. But, of course, things get a bit muddy when we consider that neutral particles have magnetic moments. One could try to justify this in most cases by noting that most such particles are either not fundamental (i.e. they are composed of other particles which \textit{aren't} neutral) or they are a direct consequence of the theory in some other way (e.g. a classical model of the photon~\cite{Brady:2015aa}). But this suggests we could never hope to develop a classical theory of the neutrino which is known to have a measurable magnetic moment~\cite{Kim:1976fk,Lee:1977kx,Marciano:1977uq,Fujikawa:1980vn,Bell:2005ys}. At the very least, it suggests that Maxwell's electromagnetic theory is incomplete.

One response to this is to simply dismiss the neutrino as non-classical and thus not subject to the rules of classical electromagnetic theory. As a response to the subject-object distinction, this sort of thinking seems evasive at best. In addition, in his paper extending beables to the realm of quantum field theory (and thus what we might blithely call the `proper' realm of the neutrino), Bell refers to the beables of a theory as ``those elements which might correspond to elements of reality, to things which exist''~\cite{Bell:2004ac}, p. 174. One presumes that matters of reality and existence are independent of any particular theory. In other words, if beables are said to properly exist in that they are elements of reality, and they are understood to be objective properties of objects, then if a magnetic field is a beable in one theory, it ought to properly be a beable in any theory in which it appears. It doesn't seem unreasonable to then ask that the nature of such beables be consistent across theories.

For the purposes of science, the existence of certain things is taken as self-evident. I may awake tomorrow to find that I am actually a Buddhist monk living in a monastery in the Himalaya and that my life as a physicist was nothing but a dream. The logician \textit{might} rightly point out that I can't disprove that. But it doesn't help me in the here and now where, dream or not, I am a physicist. As one unnamed reviewer in \textit{Philosophical Magazine} once put it, science is the ``rational correlation of experience'' (as quoted in~\cite{Eddington:1939fk}). In order to `do' science we must have some common base from which we can build our theories. So we assume that certain elements of our collective experience simply must exist. In fact the logician in the parable does not deny the existence of the cow nor even that one side of the cow is brown. The denial is only of an inferred experience. The logician takes the phrase ``rational correlation of experience'' literally in that none of the travelers `experience' (observe) the other side of the cow. They can only rationally correlate what they directly experience. Of course that's a problem for quite a few theories. Here is where Bohr and Bell are right; the world of our direct experience is classical. 

In fact the world of our direct experience is even more limited than that. We have no direct experience of electric fields in the sense that we have no way to directly measure one. We infer their existence from measurements of a the scalar electric potential. This is curious. According to Bell, electric fields are beables in classical electromagnetic theory but scalar potentials are not. Our only experience of the beables which, to Bell represent what is `physical', i.e. that which `exists', is mediated by something Bell explicitly says is `nonphysical' and thus, one would presume, does not actually exist (at least according to Bell).

Regardless of the physicality of scalar potentials we still have no known way of directly measuring an electric field. We must infer its existence from measurements of other properties. This is actually true of \textit{any} field. We cannot measure a gravitational field directly either. We infer its existence from measurements of force, acceleration, mass, etc. Bell at least partially acknowledges this fact by noting that all physical theories are necessarily tentative in nature.
\begin{quote}
Such a theory is at best a \textit{candidate} for the description of nature. Terms like `being', `beer'\footnote{To be read as \textit{be-er} not \textit{beer} (i.e. not the beverage).}, `existent', etc., would seem to me lacking in humility. In fact `beable' is short for `maybe-able'~\cite{Bell:2004ac}, p. 174.
\end{quote}
Bell also recognizes that our fundamental window on the world is through observables, but he says that our observables must be constructed from beables.\footnote{``Observables are \textit{made} out of beables''~\cite{Bell:2004aa}, p. 41.} Thus Bell acknowledges that at least some beables must be inferred. Certainly the settings of switches and knobs, and the readings of instruments, which Bell also considers beables~\cite{Bell:2004ab}, may be experienced directly. But at least some beables simply cannot be directly known.

The problem of induction, then, is in knowing just how much we can reasonably and rationally infer from a set of sensory data that constitute our direct experience of the world. In a sense, the problem of induction is concerned with just how we identify what actually is fundamental. After all, one assumes that there is some minimum set of beables required for any final theory should such a theory even be attainable. To put it another way, one assumes that the universe, at its most fundamental level, consists only of those beables that are necessary to reproduce its manifest phenomena, i.e. there should be no extraneous beables. Are these fundamental beables knowable and, if so, how can we know them?

\section{Can a universe be a `beable'?}

One approach to the problem of induction is to build theories from the ground up. That is, rather than construct theories based on our observations of the world, we could attempt to deduce them from first principles, i.e. axiomatize them. Proponents of such methods include Karl Popper~\cite{Popper:1959aa}, David Hilbert~\cite{Hilbert:1900ly,Hilbert:1901ve,Hilbert:1902zr}, and Arthur Eddington~\cite{Eddington:1936aa,Eddington:1946aa}. Eddington classified all knowledge of the physical universe as being either \emph{a priori} or \emph{a posteriori}~\cite{Eddington:1939fk}. Knowledge that is a result of a measurement (or observation) is \emph{a posteriori} while knowledge derived from an epistemological study of the actual procedure of measurement is said to be \emph{a priori}. There is a certain sub-class of such methods, that I will refer to as \textit{reductio-deductivist}, that are concerned with the minimum a priori knowledge necessary to cogently describe the fundamental aspects of the physical universe, i.e. (in some sense) its base `axioms'. In other words, beginning with the simplest axioms we can imagine, how much of the universe of our experience can we recover?

I wish to put emphasis on the phrase `universe of our experience' here. Whatever our motivations as scientists may be, we're all ultimately trying to understand the world around us. So when physicists postulate things that are far removed from everyday experience like strings or alternate universes, they are not merely engaging in mental gymnastics. Ostensibly they do so in an effort to better explain the universe of their experience. For example, a cosmologist may spend time studying a de Sitter universe, even though it is quite clear that we do not live in one, in order to better understand the universe we \textit{do} live in.

The concept of a universe would seem to present a problem for Bell's theory of local beables. As Bell himself said, ``When the `system' in question is the whole world where is the `measurer' to be found?''~\cite{Bell:2004ad}, p. 117. Perhaps it is because of this sentiment that he never seems to have considered whether or not a universe can be a beable. While it may be commonly thought that his theory was confined to \textit{local} beables, he did actually consider \textit{non}local beables as well. So whether a universe can be a beable seems to be a question worth asking, particularly if we were to choose to approach the problem of induction via reductio-deductivism, i.e. if we were to attempt to construct a universe from the ground up. What are the beables in a de Sitter universe, for instance? If we are to attempt to build a universe from the ground up, shouldn't we know what it is we are attempting to build?

The problem is that defining a universe turns out to be a tricker proposition than it might initially appear. Colloquially, a universe is defined as the totality of everything that exists~\cite{:2003uq}. The problems with this definition are numerous. First, it is not clear how a universe would be defined within the context of any theory that admits multiple universes, particularly in such a manner that they could be distinguished in some meaningful way. The nature of what we mean by a universe in such instances remains largely unsettled~\cite{Albert:1995uq,Wallace:2002kx}. Second, it is inherently ambiguous in regard to both `totality' and `existence.' The totality of all that exists to a proponent of an Everett-De Witt multiverse, for example, includes an infinite number of universes. This definition is simply too vague to qualify as a beable for any realistic theory.

Alternatively, an operationalist might define a universe as the totality of all that can be \emph{measured}. Wheeler's participatory universe takes this idea to its logical extreme by suggesting that \textit{only} things that can be measured can exist~\cite{Wheeler:1988dq}. This, of course, won't get us very far in the context of beables. In Bell's conception, the observables corresponding to measurements are constructed from beables. This implies that full knowledge of certain beables may not be possible. If our \textit{knowledge} of the world is limited to observables and observables are built from beables, it is not inconceivable to imagine that there are aspects of beables we won't---and possibly can't---ever know. It leaves open the possibility that there might be more to the world than merely what we can measure which means an operationalist definition is not well-suited for our purposes either.

Eddington noted that physical knowledge takes the form of a description of a `world' and thus \emph{defined} the universe to be this world~\cite{Eddington:1939fk}. In other words, he defined the universe to be the totality of extent of physical knowledge, i.e. ``the theme of a specified body of knowledge''~\cite{Eddington:1939fk}, p.3.  At first glance, this would appear to be very similar to the operational definition and would thus pose similar problems in relation to Bell's concept of beables. But this is only true if physical knowledge is limited by what can be directly measured. It leaves the door open to knowledge that cannot be directly measured but might possibly be reliably \emph{inferred}. But that, of course, brings us back, once again, to the problem of induction. So while Eddington's definition of the universe may not pose a direct problem for the concept of beables, it \textit{does} run into the problem of induction.

One could also attempt to define a universe topologically as some kind of spacetime manifold, but this presents at least three problems. First, it assumes that the manifold itself is somehow `real' and not merely a mathematical abstraction, e.g. a universe entirely devoid of anything---matter, fields, et. al.---would, \emph{by definition}, still have a metric. Yet it seems nonsensical to even speak of a metric for a perfectly isolated space devoid of literally anything. What meaning would space and time even have in this case? In any case, debate over the ontological status of spacetime is still ongoing~\cite{:2006vn,:2008ys}. The second problem here is that a topologically defined universe does not seem to explain emergent spacetimes (for examples of theories that involve an emergent spacetime, see~\cite{Seiberg:2005uq,Weinfurtner:2007kx,Hamma:2010vn}). Many theories that define the universe topologically do not include a mechanism for the creation of the topology in the first place (though some do). It simply is. Neither of these problems necessarily make this definition unsuitable for use as a beable. But, a topological definition of a universe seems to miss much of the detail of what is contained within it. As it happens, there is a more fundamental definition of a universe that includes the topology as well as much more.

Some theories define the universe based on a wavefunction of some kind \cite{Mersini-Houghton:2006fk}, e.g. as a solution to the Wheeler-DeWitt equation, $H\ket{\psi}=0$. One might immediately criticize this definition on the grounds that it involves a wavefunction which carries a great deal of interpretational baggage. However, if one derives the Wheeler-DeWitt equation from something like the ADM formalism~\cite{Arnowitt:1959aa}, it becomes clear that it is more formally a \textit{field} equation. Solutions to the Wheeler-DeWitt equation are not spatial wavefunctions in the sense implied by non-relativistic quantum mechanics. Rather they are functionals of field configurations taken on all of spacetime. As such, the Hamiltonian, though still an operator in a Hilbert space that acts on wavefunctions, is not quite the same beast as in non-relativistic quantum mechanics. The solutions to the Wheeler-DeWitt equation, which are still wavefunctions (just not in a spatial sense), contain all the information concerning the matter and geometry of the universe, i.e. the topology of spacetime and the matter therein. So, in that sense, solutions to the Wheeler-DeWitt equation would seem to be a more fundamental definition of a universe than one based solely on a topology. 

Wavefunctions do not necessarily pose a problem for Bell's concept of beables. Though it is commonly thought that his theory was one consisting exclusively of \textit{local} beables, i.e. ones confined to a particular spacetime region, which would seem to rule out wavefunctions, this is, in fact, not entirely true. In~\cite{Bell:2004ac} he makes the point that it is essential that any theory be able to define the positions of things including instrument pointers since these tell us the results of measurements. In attempting to make the idea `positions of things' more precise, he chooses to use the fermion number density since the distribution of fermion number in the universe should include the `positions of things' (and a great deal more). But he then goes on to say that
\begin{quote}
[t]he lattice fermion number are the local beables of the theory, being associated with definite positions in space. The state vector $\ket{t}$ also we consider as a beable, although \textit{not a local one}~\cite{Bell:2004ac}, p. 176. [my emphasis]
\end{quote}
So he grants beable status to the state vector. The state vector $\ket{t}$ evolves in time according to the Schr\"{o}dinger equation and the usual Hamiltonian operator. Wavefunctions can, of course, be constructed from state vectors, though that does not necessarily make them beables. Remember that observables are constructed from beables and can occasionally be promoted to the status of beable, but are not usually beables themselves. In fact, it is worth noting that in~\cite{Bell:2004ab}, p. 53, he explicitly does \textit{not} grant beable status to the usual, spatial wavefunction due to the nonlocality associated with its instantaneous collapse over all space upon measurement. But the wavefunction in the Wheeler-DeWitt equation suffers from no such defect since it is said to be `timeless', i.e. it simply `is'.

It is worth pausing here and briefly reviewing the nature of beables. Bell's definition of the term actually includes subtle variations over the many publications in which he employs it. Likely these represent an evolution of his thinking on the subject. One initially gets the impression that beables must be classical things such as pointers and knobs and instruments and, perhaps, fields (as long as they are classical or, in his words, `physical'). Later, Bell suggests that beables are what `exist.' In his discussion of beables in quantum field theory, he leaves any classical notion behind, granting beable status to fermion number density and the state vector. Nevertheless, the concept of `beable' is very clearly meant to define the ontology of a theory, i.e. what the theory is about.

So where does that leave us? If we define a universe as a solution or set of solutions to the Wheeler-DeWitt equation, where those solutions are functionals of field configurations, then it seems that a universe \textit{can} be a beable and can thus serve as an ontology for a theory. But is there something more that can be said?

\section{Is the universe fundamental?}

Beables serve as the ontology of theories but are they fundamental? In other words, is there a reality deeper than beables? This is an interesting question in the context of a universe. If we were to na\"{i}vely think of a universe as an object and then ask whether or not it was fundamental, the answer would be unclear since we know that most universes contain things like matter and energy and we might consider our universe to be constituted \textit{of} such things. But that's not the sense I mean in this instance.

To Bell, beables certainly are fundamental within a given theory since they form the ontology for that theory, i.e. what the theory is about. For most theories, we can think of the universe as a bit like the substrate on which the beables of those theories reside. But for theories concerning the universe itself, we want to know more about the substrate. To put it another way, if we were to build a universe from scratch in a reductio-deductivist sense, it seems logical to start by formulating a wavefunction as a functional of a set of field configurations. But there's a problem with this. Presumably, if we have defined our universe in terms of a functional of some set of field configurations, one would assume that \textit{we would also need to define the fields.} Our definition of a universe \textit{references something else.} That suggests that, at least in this context, we have a beable for a theory that is \textit{not} actually fundamental. Of course there is nothing inherently wrong with this in the sense of Bell's conception of beables since he made no explicit requirement that they be fundamental. On the other hand, how could a universe \textit{not} be fundamental? It's hard to imagine anything more fundamental than a universe.

But, let's return for a moment to the colloquial definition of a universe as the totality of all that exists. In a way, that definition suggests that the concept of a universe is meaningless without something else, specifically \textit{all that exists}. In that sense, a universe \textit{isn't} fundamental. What actually defines it is that from which it is constructed. A universe without structure, without elements is meaningless. As such, a `universe', as envisaged here and consistent with Bell's notion of beables, \textit{is not fundamental}.

We shouldn't read \textit{too} much into this conclusion, however. This result applies only to universes that can be modeled using Bell's notion of beables. It is possible that there might be ways to define a universe that cannot be a beable in any theory. In addition, given the minor ambiguities associated with Bell's notion of beables and the way in which the idea took shape in his writing over the years, it is possible to reach a different conclusion in this matter. But it is important to remember that the wavefunction in the Wheeler-DeWitt equation is not the same sort of thing as the wavefunction in non-relativistic quantum mechanics. It doesn't suffer from the same nonlocal transformation. In fact it doesn't transform at all! It simply is. As Bell said when he introduced the concept of beables, ``it should again become possible to say of a system not that such and such may be \textit{observed} to be so but that such and such \textit{be} so''~\cite{Bell:2004aa}. The universe's existence is independent of our observation of it. We don't simply \textit{observe} that it exists, it \textit{does} exist. Of that I am sure, even if I wake up tomorrow in a monastery.

\begin{acknowledgements}
I would like to thank Travis Norsen, Tim Maudlin, Hans Westman, and Travis Myers for a stimulating and enlightening discussion that helped to shape this essay.
\end{acknowledgements} 

\newpage

\bibliographystyle{plain}
\bibliography{FQXi7.bib}

\end{document}